# Symbolic Synthesis of Knowledge-based Program Implementations with Synchronous Semantics

X. Huang
xiaoweih@cse.unsw.edu.au

R. van der Meyden
meyden@cse.unsw.edu.au

## ABSTRACT

This paper deals with the automated synthesis of implementations of knowledge-based programs with respect to two synchronous semantics (clock and synchronous perfect recall). An approach to the synthesis problem based on the use of symbolic representations is described. The method has been implemented as an extension to the model checker MCK. Two applications of the implemented synthesis system are presented: the muddy children puzzle (where performance is compared to an explicit state method for a related problem implemented in the model checker DEMO), and a knowledge-based program for a dynamic leader election problem in a ring of processes.

## Categories and Subject Descriptors

D.2.4 [**Software Engineering**]: Software/Program Verification; F.4.1 [**Mathematical Logic**]: Modal Logic

## 1. INTRODUCTION

One of the main motivations for the application of epistemic logic in computer science has been the observation that it provides a beneficial level of abstraction through which to view distributed systems. A range of problems in distributed computing have been studied from this perspective, including protocols for agreement [17, 8, 9, 27], message transmission [19], atomic commitment [15, 25], clock synchronization [26, 28], leader election [18], and secure communication [35, 1, 2, 24, 36].

Many of these analyses are based on first expressing the solution to a problem in terms of the relation between an agent's actions and its knowledge, and then seeking to understand the conditions under which the agent has the requisite knowledge. A first codification of the approach was the semantic notion of *knowledge-based protocols* of [16], and the idea was refined and given a syntactic basis in the *knowledge-based programs* of [11, 10]. The latter provide a simple guarded command programming notation (in the style of Unity [7]), in which the guards in conditional statements are not just expressions over the agent's local variables, but may also contain formulas of epistemic logic asserting some property of the agent's knowledge.

Knowledge-based programs resemble standard proagrams, but they do not have a straightforward operational semantics. Instead, they are semantically more like a specification, in that they stand in an *implementation* relation to standard programs. To obtain an implementation, one must replace the knowledge conditions in the program by expressions in the agents' local variables that are equivalent, when running the resulting standard program. Because of the *fixpoint* nature of this semantics, in general, a knowledge-based program could have no, one, or many behaviourally distinct implementations. There are, however, some syntactic and semantic conditions under which implementations are guaranteed to be unique [11]. One of these is that the formulas appearing in the knowledge conditions are free of temporal operators and that the semantics of the knowledge operators is *synchronous*, in the sense that agents always know the current time.

The early literature on knowledge in distributed computing and knowledge-based programs is confined to "pencil and paper" analyses. In recent years, automated tool support for knowledge-based analysis has begun to be developed, in the form of *epistemic model checkers* [13, 23, 30, 20, 37], which are able to automatically verify whether standard programs satisfy epistemic specifications. These model checkers have been applied to a number of case studies in which it is verified that a proposed implementation of a knowledge-based program is indeed an implementation [3, 2, 24]. However, the approach applied in these studies still requires that the proposed implementation be derived manually. Since the implementations may make use of subtle sources of information, this can be a highly nontrivial task, although the examples automatically constructed by the model checker when checking an incorrect implementation can provide useful information to guide the search [3, 1].

Our contribution in this paper is to develop the first practical tool for automated synthesis of knowledge-based program implementations, by extending methods from epistemic model checking. The main contributions of the paper are as follows:

1. We develop a practical syntax for knowledge-based programs that extends the Unity style programs of [11] to encompass use of knowledge in assignment statements, as well as sequential structure.

2. We show how existing symbolic techniques for epistemic model checking may be extended to yield an approach to automated synthesis of knowledge-based program implementations. Our techniques work for the special case of atemporal knowledge-based programs with respect to two distinct synchronous semantics for knowledge, the *clock* and *synchronous perfect recall semantics*, in which, as noted above, unique implementations are guaranteed to exist.

3. We have implemented these algorithms as an extension of the epistemic model checker MCK. One benefit of building on the existing model checking technology is that properties of the implementation derived can directly be verified, with many of the computational steps required for verification already performed by the synthesis procedure.

4. We conduct a number of validation case studies of knowledge-based program implementation using the resulting tool, considering two types of examples. In the first, the muddy children puzzle, we compare the performance of our symbolic synthesis approach to the performance of the model checker DEMO. DEMO does not synthesize implementations of knowledge-based programs, but solves a closely related model checking problem using an explicit state rather than symbolic technique. The second example we consider is a knowledge-based program for a leader election protocol in a ring of processes.

The structure of the paper is as follows. In Section 2, we review the basics of epistemic model checking and a symbolic technique used in the implementation of such systems. We develop a syntax and semantics for knowledge-based programs in Section 3. Section 4 describes the basis for a symbolically implementable procedure for synthesis of knowledge-based program implementations. The application of this procedure to a number of examples is discussed in Section 5. Section 6 discusses related work, and we make some concluding remarks in Section 7.

## 2. EPISTEMIC MODEL CHECKING

In this section, we recall the background we require from epistemic logic (following [10]) and epistemic model checking (following [34]).

Let $Prop$ be a set of atomic proposition and let $Ags$ be a finite set of agents. The temporal-epistemic logic that we work with has the syntax

$$\phi ::= p \mid \neg\phi \mid \phi_1 \wedge \phi_2 \mid X\phi \mid K_i\phi$$

where $p \in Prop$ and $i \in Ags$. Intuitively, formula $X\phi$ expresses that $\phi$ holds at the next time, and $K_i\phi$ expresses that agent $i$ knows that $\phi$ holds. A formula is *atemporal* if it does not make use of the temporal operator $X$.

At all times, each agent $i$ is assumed to be in some local state that records all the information that it can access at that time. The environment $e$ records "everything else that is relevant". Let $S$ be the set of environment states and let $L_i$ be the set of local states of agent $i$. A *global state* $s$ of a multi-agent system is a tuple $(s_e, s_1, ..., s_n)$ such that $s_e \in S$ and $s_i \in L_i$ for all $i \in Ags$.

A *run* $r$ is a function from time to global states, i.e., $r : \mathbb{N} \to S \times L_1 \times ... \times L_n$. A pair $(r, m)$ consisting of a run $r$ and a time $m$ is called a point. A *system* $\mathcal{R}$ is a set of runs. We call $\mathcal{R} \times \mathbb{N}$ the set of points of $\mathcal{R}$. If $r(m) = (s_e, s_1, .., s_n)$ then for $x \in Ags \cup \{e\}$ we write $r_x(m)$ for $s_x$ and $r_x(0..m)$ for $r_x(0)...r_x(m)$. Relative to a system $\mathcal{R}$, we define the set $\mathcal{K}_i(r,m) = \{(r', m') \in \mathcal{R} \times \mathbb{N} \mid r_i(m) = r'_i(m')\}$ to be the set of points that are indistinguishable from the point $(r, m)$ for agent $i$.

An interpreted system $\mathcal{I}$ is a tuple $(\mathcal{R}, \pi)$ such that $\mathcal{R}$ is a system and $\pi : \mathcal{R} \times \mathbb{N} \to \mathcal{P}(Prop)$ is an assignment giving an interpretation to the atomic propositions at each point. Given an interpreted system $\mathcal{I}$, a point $(r, m)$, and a formula $\phi$, we define the relation $\mathcal{I}, (r, m) \models \phi$ inductively by

- $\mathcal{I}, (r, m) \models p$ if $p \in \pi(r, m)$
- $\mathcal{I}, (r, m) \models \neg\phi$ if not $\mathcal{I}, (r, m) \models \phi$
- $\mathcal{I}, (r, m) \models \phi_1 \wedge \phi_2$ if $\mathcal{I}, (r, m) \models \phi_1$ and $\mathcal{I}, (r, m) \models \phi_2$
- $\mathcal{I}, (r, m) \models X\phi$ if $\mathcal{I}, (r, m+1) \models \phi$
- $\mathcal{I}, (r, m) \models K_i\phi$ if $\mathcal{I}, (r', m') \models \phi$ for all points $(r', m') \in \mathcal{K}_i(r, m)$

Since interpreted systems are infinite structures and for model checking we require a finite input, we generate interpreted systems from finite structures. A *(finite state) transition model* $M$ for agents $Ags$ is a tuple $M = (S, I, \{O_i\}_{i \in Ags}, \to, \pi)$, where $S$ is a (finite) set of states, $I \subseteq S$ is the set of initial states, each $O_i : S \to O$ is a function representing the observation that agent $i$ makes at each state, $\to \subseteq S^2$ is a serial transition relation over states in $S$, and $\pi : S \to \mathcal{P}(Prop)$ is a propositional assignment. Let $k_i(s) = \{s' \in S \mid O_i(s) = O_i(s')\}$ be the set of states that are indistinguishable from state $s$ for agent $i$, based on its observation.

A *path* $\rho$ from a state $s$ of $M$ is a finite or infinite sequence of states $s_0 s_1 ...$, such that $s_0 = s$ and $s_k \to s_{k+1}$ for all $k < |\rho| - 1$, where $|\rho|$ is the total number of states in $\rho$. Given such a path $\rho$, we use $\rho(m)$ to denote the state $s_m$. A *fullpath* from a state $s$ is an infinite path from $s$. A *path* $\rho$ is initialized if $\rho(0) \in I$.

To obtain a system from a finite state transition model $M$, we treat the states of $M$ as the states of the environment, and obtain runs from paths by adding local states at each point. This can be done in a variety of ways, representing different levels to which agents recall their observations. We call the level of recall a *view* and deal with the views obs, clk and spr representing recall only of the current observation, recall of the current observation and the time and synchronous perfect recall, respectively.

For each initialized fullpath $\rho$ and view $\mathcal{V} \in \{\text{obs}, \text{clk}, \text{spr}\}$, we define a run $\rho^\mathcal{V}$. The state of the environment at time $m$ is given by $\rho_e^\mathcal{V}(m) = \rho(m)$ in each case, and the agents' local states are assigned as follows:

- $\mathcal{V} = \text{obs}$: the local state of agent $i$ at time $m$ is $\rho_i^{\text{obs}}(m) = O_i(\rho(m))$;
- $\mathcal{V} = \text{clk}$: the local state of agent $i$ at time $m$ is $\rho_i^{\text{clk}}(m) = (m, O_i(\rho(m)))$;
- $\mathcal{V} = \text{spr}$: the local state of agent $i$ at time $m$ is $\rho_i^{\text{spr}}(m) = O_i(\rho(0))...O_i(\rho(m))$.

Given a system $M$ and a view $\mathcal{V}$, we write $\mathcal{R}^\mathcal{V}(M)$ for the set of runs $\rho^\mathcal{V}$ where $\rho$ is an initialized full-path of $M$. The interpretation $\pi$ of $M$ lifts to an interpretation $\pi^\mathcal{V}$ on the global states in $\mathcal{R}^\mathcal{V}(M)$, defined by $\pi^\mathcal{V}((s, l_1, \ldots, l_n)) = \pi(s)$. We define the interpreted system obtained from $M$ using view $\mathcal{V}$ by $\mathcal{I}^\mathcal{V}(M) = (\mathcal{R}^\mathcal{V}(M), \pi^\mathcal{V})$. Given a finite model $M$, a view $\mathcal{V}$, and a formula $\phi$, we write $M \models^\mathcal{V} \phi$ if $\mathcal{I}^\mathcal{V}(M), (r, 0) \models \phi$ for all $r \in \mathcal{R}^\mathcal{V}(M)$.

The *model checking problem* is to determine, given a finite state transition model $M$, a view $\mathcal{V}$ and a temporal epistemic formula $\phi$, whether $\mathcal{I}^\mathcal{V}(M) \models \phi$. *Epistemic model checkers* are software systems that solve this problem. A number of such systems have been implemented. MCK [13] supports all three views, MCMAS [23], Verics [20] and MCTK [30] work with the observational view. These systems use a variety of temporal logics for the temporal expressiveness in formulas. MCK supports a superset of the language defined above.

Before concluding this section, we define a presentation of standard $S5_n$ Kripke-structures that will be used later. An *epistemic model* is a tuple $\mathtt{M} = (S, \{O_i\}_{i \in Ags}, \pi)$, where the components are of the same type as the similarly named components in state transition models. Given a model $\mathtt{M}$, a state $s \in S$, and an *atemporal* formula $\phi$, the relation $\mathtt{M}, s \models \phi$ can be recursively defined as follows:

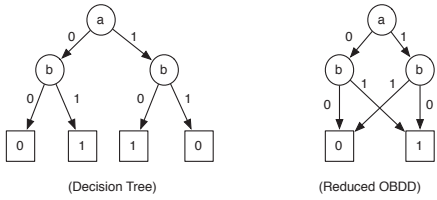

Figure 1: A decision tree and its reduced OBDD

- $M, s \models p$ if $p \in \pi(s)$

- $M, s \models \neg \phi$ if not $M, s \models \phi$

- $M, s \models \phi_1 \wedge \phi_2$ if $M, s \models \phi_1$ and $M, s \models \phi_2$

- $M, s \models K_i \phi$ if $M, s' \models \phi$ for all states $s'$ such that $O_i(s) = O_i(s')$

It is easily seen that for atemporal formulas $\psi$ that are boolean combinations of formulas of the form $K_i \phi$ (for a fixed $i$), for all $s, s' \in S$ with $O_i(s) = O_i(s')$ we have $M, s \models \psi$ iff $M, s' \models \psi$, i.e., the truth value of $\psi$ depends only on $O_i(s)$. For $o \in O_i(S)$, we may therefore define the relation $M, o \models_i \psi$ if $M, s \models \psi$ for some (equivalently, all) $s \in S$ with $O_i(s) = o$.

## 2.1 Symbolic Data Structures

MCK supports a number of different algorithmic approaches to solving the epistemic model checking problem. One of these is based on symbolic model checking using (reduced) ordered binary decision diagrams (BDD) [6]. These are data structures defined as follows.

Let $V$ be a set of variables. A *V-assignment* is a function $s : V \rightarrow \{0, 1\}$. Write $Assgts(V)$ for the set of all $V$-assignments, and $s[v \mapsto x]$ for the function that is identical to $s$ except that it takes value $x$ on input $v$. A *V-indexed boolean function* is a mapping $f : Assgts(V) \rightarrow \{0, 1\}$. Note that such functions are able to represent sets $X \subseteq Assgts(V)$ by their characteristic functions $f_X$, mapping $s$ to 1 just in case $s \in X$. One way to represent such a function $f$ is using a *binary tree* of height $n$, with each level corresponding to one of the variables in $V$, and leaves labelled from $\{0, 1\}$. This tree can in turn be thought of as a finite state automaton on alphabet $\{0, 1\}$. Reduced ordered binary decision diagrams (BDD's in the sequel) more compactly represent such a function as a *dag* of height $n$, with binary branching, by applying the usual finite state automaton minimization algorithm. A very simple example of this for the function $f(a, b, c) = a$ xor $b$ is illustrated in Figure 1. In some cases, the degree of compaction obtained in the minimal dag representation is considerable. We note that the amount of compaction obtained is sensitive to the variable ordering used, and finding a variable ordering that minimizes the result is NP-hard, though there exist good heuristics, such as *sifting* [29].

Given this minimal representation of $V$-indexed boolean functions, it is moreover possible to compute (in practice, often in reasonable time) some operations on these functions, by means of algorithms that take as input the BDD representation of the input functions and returns the BDD representation of the result. The operations for which this can be done include the following:

- Boolean operations $\wedge$, $\neg$, defined pointwise on functions. E.g., if $f, g : Assgts(V) \rightarrow \{0, 1\}$, then $f \wedge g : Assgts(V) \rightarrow \{0, 1\}$ is defined by $(f \wedge g)(s) = f(s) \wedge g(s)$.

- Boolean quantification $\exists, \forall$, e.g., if $f : Assgts(V) \rightarrow \{0, 1\}$ and $v \in V$ then $\exists v(f) : Assgts(V \setminus \{v\}) \rightarrow \{0, 1\}$ maps $s \in Assgts(V \setminus \{v\})$ to $f(s[v \mapsto 0]) \vee f(s[v \mapsto 1])$.

- variable substitution: if $f : Assgts(V) \rightarrow \{0, 1\}$ and $U \subseteq V$ and $U'$ are sets with $U' \cap (V \setminus U) = \emptyset$, and $\sigma : U \rightarrow U'$ is a bijection, then $f_\sigma : Assgts((V \setminus U) \cup U') \rightarrow \{0, 1\}$ maps $s : Assgts((V \setminus U) \cup U')$ to $s'$, where $s'(v)$ is $s(v)$ when $v \in V \setminus U$ and $s(\sigma^{-1}(v))$ when $v \in U'$.

*Symbolic* model checking, as implemented in MCK, then proceeds using BDD representations of sets and relations relevant to model checking. For example, the set $I$ of initial states of a system can be represented as a BDD-encoded boolean function $f_I$ indexed by the state variables $V$. Relations can be represented using "primed" versions of the state variables $V$, defined by $V' = \{v' \mid v \in V\}$. A relation such as the transition relation $\rightarrow$ of a model can then be represented as a function $f_\rightarrow$ indexed by variables $V \cup V'$, such that if $s$ and $t$ are assignments to $V$, we have $s \rightarrow t$ iff $f_\rightarrow(s \cup t') = 1$, where $t'$ is obtained from $t$ by renaming each variable $v$ to its primed counterpart $v'$. Operations such as the composition of a relation and a set can then be performed at the level of the BDD representation, e.g. $\{t \in S \mid s \in I \wedge s \rightarrow t\}$ is represented by the function $\exists V(f_I \times 1_{V'} \wedge f_\rightarrow)\sigma$ on $V$, where $f_I \times 1_{V'}$ trivially extends $f_I$ by adding (irrelevant) variables $V'$, and $\sigma$ renames the variables $V'$ to the variables $V$ by removing the prime symbol.

## 3. KNOWLEDGE-BASED PROGRAMS

Knowledge-based analyses of systems typically concern the interaction between agents' knowledge and their actions. *Knowledge-based programs* [11, 10] have been proposed to capture such relationships in a program-like notation, with actions chosen according to conditions expressed in epistemic logic.

The original presentation of knowledge-based programs used a very simplified (Unity style [7]) programming notation, consisting of a single infinitely repeated do loop containing a set of guarded statements of the form $\phi \rightarrow a$ where $\phi$ is an epistemic formula and $a$ an action. We develop here a slightly richer and more structured notation, using sequential composition and an epistemic assignment statement. The notation is based on the modelling notation already employed by MCK. We focus on *atemporal* programs with a synchronous semantics for knowledge (either the clock or synchronous perfect recall semantics), since this is a case in which unique implementations are guaranteed to exist.

Since, in general, even atemporal knowledge-based programs may not have finite state implementations under the perfect recall semantics, we also limit ourselves to terminating programs, so omit looping from the language. Our handling of parallelism and actions (signals) is somewhat in the spirit of synchronous languages such as Esterel [5]. To give the semantics of knowledge-based programs, we use a formulation based on [32], which allows flexibility in choice of view based on a notion of *environment* that replaces the notion of *context* of [11, 10].

## 3.1 Standard Programs: Syntax

Define a *standard program* over a set $V$ of variables and a set $A$ of atomic statements to be either the terminated program $\epsilon$ or a sequence $P$ of the form $stat_1 ; \ldots ; stat_k$, where the $stat_i$ are simple statements and ';' denotes sequential composition. Each simple statement $stat_i$ is either an atomic statement in $A$ or a nondeterministic branching statement of the form

$$\text{if } g_1 \rightarrow a_1 \;[]\; g_2 \rightarrow a_2 \;[] \ldots [] \; g_k \rightarrow a_k \text{ fi}$$

where each $a_i$ is an atomic statement in $A$ and the $g_i$ are boolean expressions over $V$ called *guards*. Intuitively, a nondeterministic branching statement executes by performing one of the assignments $a_i$ for which the corresponding guard $g_i$ is true. If several guards hold simultaneously, one of the corresponding actions is selected nondeterministically. We treat $P$ as identical to $P; \epsilon$. The *length* of a program is the number of simple statements it contains. We use standard programs to describe both the behavior of agents and the environment in which they operate. The type of atomic statements used in these two cases is different.

Environment models are used to represent how states of the environment are affected by actions of the agents. Formally, we define an *environment model* to be a tuple $\mathcal{M}_e = (Ags, Acts, Var_e, Init_e, \tau)$ where $Ags$ is a set of agents, $Acts$ is a set of actions available to the agents, $Var_e$ is a set of (boolean) environment variables[1], $Init_e$ is an initial condition, in the form of a boolean formula over $Var_e$, and $\tau$ is a *transitions clause* for the environment $e$, expressed in the form of a standard program.

In addition to the environment variables $Var_e$, an additional set $ActVar(\mathcal{M}_e) = \{i.a \mid i \in Ags,\ a \in Acts\}$ of (boolean) *action variables* are generated for each model $\mathcal{M}_e$. Intuitively, $i.a$ represents that agent $i$ performs action $a$. The transitions clause is a standard program over the set of variables in $Var_e \cup ActVar(\mathcal{M}_e)$ and the set of atomic actions of the form $x := expr$, where $x \in Var_e$ and $expr$ is a boolean expresssion over $Var_e \cup ActVar(\mathcal{M}_e)$. The statement $x := expr$ represents that the value of the expression $expr$, is assigned to the variable $x$.

Protocols are used to describe the behaviour of the agents. A *protocol for agent $i$ in environment model $\mathcal{M}_e$ (of length $m$)* is a tuple $Prot_i = (PVar_i, LVar_i, OVar_i, Init_i, Acts_i, Prog_i)$ where $PVar_i \subseteq Var_e$ is a set of *parameter variables*[2], $LVar_i$ is a set of *local variables*, $OVar_i \subseteq PVar_i \cup LVar_i$ is a set of *observable variables*, $Init_i$ is an initial condition, in the form of a formula over $LVar_i$, and $Prog_i$ is standard program of length $m$. The guards in $Prog_i$ are over the set of variables $PVar_i \cup LVar_i$. The atomic statements in $P$ have the form

$$\ll a \mid x_1 := expr_1, ..., x_m := expr_m \gg$$

where $a \in Acts \cup \{nil\}$ and each $x_1 := e_i$ is an assignment statement with $x_i$ in $LVar_i$ and $e_i$ an expression over $PVar_i \cup LVar_i$. Intuitively, such an atomic statement is executed by emitting action $a$ as a signal to the environment: when agent $i$ performs the action, the variable $i.a$ is set to be true (and all other action variables $i.b$ set to be false.) The environment transition clause then runs to update the environment variables. Concurrently, the statement performs the *simultaneous assignment* $x_1 := expr_1, ..., x_m := expr_m$ in a single step of computation. That is, the expressions $e_i$ are first evaluated in the state from which the atomic statement is performed, and their values are then simultaneously assigned to the variables $x_i$. We abbreviate an atomic statement of the form $\ll nil \mid x := expr \gg$ to $x := expr$, and also abbreviate $\ll nil \mid \gg$ to *skip*.

A *joint protocol (of length $m$)* is a tuple **Prot** associating a protocol $\mathbf{Prot}_i$ (of length at most $m$) with each $i \in Ags$. A *system model* is a pair $\mathcal{S} = (\mathcal{M}_e, \mathbf{Prot})$ consisting of an environment model $\mathcal{M}_e$ and a joint protocol **Prot** for $\mathcal{M}_e$. This represents a set of agents running particular protocols in the context of a given environment.

---

[1] To simplify the presentation we assume here that all variables are boolean; our implementation also allows variables to have a declared finite type.

[2] In the concrete MCK syntax these may be given using new variables in a parameter declaration as aliases for environment variables. This allows sharing of protocol code between agents running similar programs but with different parameter bindings; see the examples below.

## 3.2 Standard Programs: Semantics

We now show how a system model generates a finite state transition model. To do so, we first convert the system model into a simple form of parallel program and provide these programs with an operational semantics.

We assume we are given a system model $\mathcal{S} = (\mathcal{M}_e, \mathbf{Prot})$, where $\mathcal{M}_e = (Ags, Acts, Var_e, Init_e, \tau)$ and

$$\mathbf{Prot}_i = (PVar_i, LVar_i, OVar_i, Init_i, Acts_i, Prog_i)$$

for $i \in Ags$. We define global states with respect to this model to be boolean assignments $s$ over the set of variables $Var_e \cup \bigcup_{i \in Ags} LVar_i$. We also define the *parallel program*

$$Prog(\mathcal{S}) = \tau \parallel_{i \in Ags} Prog_i .$$

This is an expression representing $|Ags| + 1$ components, i.e., the specially identified environment component $\tau$, a program over environment variables, and the $|Ags|$ components $Prog_i$, representing programs associated to the agents.

Intuitively, the operational semantics of these parallel statements defines a transition relation on global states. The definition of the transition relation is given in three stages, captured in the following rule:

$$\frac{\bigwedge_{i \in Ags}\ (s, P_i) \hookrightarrow\ \ll \mathbf{a}_i \mid \alpha_i \gg; P'_i,\quad (s \cup \mathbf{a}, \tau) \longrightarrow^* (s' \cup \mathbf{a}, \epsilon),}{(s, \tau \parallel_{i \in Ags} P_i) \to (s'\theta, \tau \parallel_i P'_i)}$$
$$\theta = \{x \mapsto e(s) \mid \text{``}x := e\text{''} \in \alpha_i,\ i \in Ags\}$$

The explanation of this statement is as follows: in the first stage, given a global state $s$, with its remaining computation represented by program $P_i$, each agent $i$ first generates an atomic statement $a = \ll \mathbf{a}_i \mid \alpha_i \gg$ as well as a program $P'_i$, to be run after this atomic statement has executed. This is represented formally by a relation $(s, P_i) \hookrightarrow a; P'_i$. In the second stage, the actions in these statements are then combined into a joint atomic action $\mathbf{a}$, viewed as an assignment making the action variables $i.\mathbf{a}_i$ true for $i \in Ags$, and all other action variables false. This assignment is added to the current global state, and the environment program $\tau$ then causes a transition of the environment state, expressed by the statement $(\tau, s \cup \mathbf{a}) \longrightarrow^* (s' \cup \mathbf{a}, \epsilon)$ that represents that the environment program $\tau$, when executed with respect to joint action $\mathbf{a}$, runs to termination having caused the global state to change from $s$ to $s'$ (only the environment variables change during the running of $\tau$). Finally, the local states are updated, by executing the assignments $\alpha_i$ locally at each agent. This is captured by first defining the substitution $\theta$ that defines the update to be performed, based on the values $e(s)$ of expressions $e$ in the state $s$, and then applying that substitution to the global state $s'$ (represented by $s'\theta$).

The relations used above are defined by the following rules:

$$\overline{(s, \epsilon) \hookrightarrow skip; \epsilon} \qquad\qquad \overline{(s, a; P) \hookrightarrow a; P}$$

$$\frac{s \models g_i}{(s, \mathbf{if}\ g_1 \to a_1\ []\ ...\ []\ g_m \to a_m\ \mathbf{fi}; P) \hookrightarrow a_i; P}$$

$$\frac{s \models \bigwedge_{i \in Ags} \neg g_i}{(s, \mathbf{if}\ g_1 \to a_1\ []\ ...\ []\ g_m \to a_m\ \mathbf{fi}; P) \hookrightarrow skip; P}$$

$$\frac{(s, P) \hookrightarrow x := e; P',\quad \theta = [x \mapsto e(s)]}{(s, P) \to (s\theta, P')}$$

(The rules for $\hookrightarrow$ apply to both the environment program and the agent protocols; the last rule applies only to steps of the environment computation.)

We can now define a model $M(\mathcal{S}) = (S, I, \{O_i\}_{i \in Ags}, \to, \pi)$ for each system model $\mathcal{S}$. The components are given as follows: $S$ is the set of pairs $(s, \tau\|_{i\in Ags}P_i)$, where $s$ is a global state of $\mathcal{S}$ and each $P_i$ is a protocol for agent $i$, the set $I$ is the set of pairs $(s, Prog(\mathcal{S}))$ such that $s \models Init_e \wedge \bigwedge_{i\in Ags} Init_i$, the function $O_i$ is defined by $O_i((s, \tau\|_{i\in Ags}P_i)) = s \upharpoonright OVar_i$, the transition relation $\rightarrow$ is as defined above, and $\pi$ associates each variable with its value, i.e. $v \in \pi(s)$ iff $s(v) = 1$.

Note that, given a view $\mathcal{V}$, we obtain from $M(\mathcal{S})$ an interpreted system $\mathcal{I}^{\mathcal{V}}(M(\mathcal{S}))$. We use this construction of interpreted systems to give semantics to knowledge based programs.

### 3.3 Knowledge-based protocols

The syntax of knowledge-based protocols is given as a generalization of the definitions above. A *knowledge-based protocol for agent i in environment* $\mathcal{M}_e$, is a tuple

$$P_i = (PVar_i, LVar_i, OVar_i, Init_i, Acts_i, \text{Prog}_i),$$

where the components are exactly as for a protocol for agent $i$ in environment $\mathcal{M}_e$, except that in the program $\text{Prog}_i$, both the guards $g$ in conditional statements and the expressions $e$ in the assignments in atomic statements may be formulas of the logic of knowledge. Figure 2(a) gives an example of such a program, (corresponding to a stage of the well-known "Muddy Children" problem, which we discuss in more detail in Section 5). A *joint knowledge-based protocol* is a tuple $P = \{P_i\}_{i\in Ags}$ consisting of a knowledge-based protocol $P_i$ for each agent $i$.

To give semantics to knowledge-based protocols, we define a relation of *implementation* between knowledge-based protocols and standard protocols. Intuitively, an implementation is a standard protocol that is structurally similar to the knowledge-based protocol, except that knowledge formulas have been replaced by expressions in the local variables, where such expressions are equivalent to the knowledge formulas. To make sense of this equivalence we need to evaluate the knowledge formulas in an interpreted system: for this we take the system generated by the standard protocol.

We first give the semantics with respect to the clock view. Note that since programs are sequences $stat_1; \ldots; stat_m$ of simple statements, each such simple statement can be associated with a time of occurrence, viz., $stat_i$ occurs at time $i - 1$. In case $stat_i$ is a conditional statement **if** $g_1 \rightarrow a_1 \;[] \ldots [] \; g_k \rightarrow a_k$ **fi** we also say that each of the atomic statements $a_j$ occur at time $i - 1$. (Intuitively, it takes no time to evaluate the guard $g_j$.) Given a knowledge-based program $\text{Prog}_i$, we transform it into its *skeleton*, denoted $skell(\text{Prog}_i)$, by replacing each knowledge formula $\phi$ in a guard $g$ or assigned expression $e$, occurring at time $t$, by a new variable $v_\phi^t$, whose name indicates both the time $t$ and the formula being replaced. (More precisely, we replace the maximal subformulas $\phi$ that contain knowledge operators but do not contain "non-observable" variables in $PVar_i \setminus OVar_i$.) Let $skellVar(\text{Prog}_i)$ be the set of such new variables in $skell(\text{Prog}_i)$. We define $skell(P) = \{skell(\text{Prog}_i)\}_{i\in Ags}$ and $skellVar(P) = \cup_{i\in Ags} skellVar(\text{Prog}_i)$.

Next, let $\theta$ be a substitution mapping each skeleton variable $v_\phi^t \in skellVar(\text{Prog}_i)$, for $i \in Ags$, to a boolean expression on the observable variables of agent $i$'s protocol $P_i$. If we apply this substitution to $skell(\text{Prog}_i)$, we obtain a standard program $skell(\text{Prog}_i)\theta$. We write $P_i\theta$ for the result of replacing the knowledge-based program $\text{Prog}_i$ in $P_i$ by $\text{Prog}_i\theta$. This is a standard protocol for agent $i$.

Similarly, if $P = \{P_i\}_{i\in Ags}$ is a *joint knowledge-based protocol*, and $\theta$ is a substitution satisfying the condition above for all agents $i$, we write $P\theta$ for the joint standard protocol $\{P_i\theta\}_{i\in Ags}$. We now define $P\theta$ to be an *implementation* of the joint knowledge-based protocol $P$ with respect to the view clk if $\mathcal{I}^{\text{clk}}(M(\mathcal{M}_e, P\theta)) \models X^t(\phi \Leftrightarrow \theta(v_\phi^t))$ for all $v_\phi^t \in skellVar(P)$. That is, in the system obtained with respect to the view clk by running the standard protocol $P\theta$ in the envi-ronment $\mathcal{M}_e$, each knowledge formula $\phi$ in $P$ is equivalent to the concrete expression $\theta(v_\phi^t)$ on the local state variables that replaces it in the standard protocol (at the time $t$ that this formula is relevant to the behaviour of the program).

Since the definition of implementation of a knowledge-based program is stated as a constraint on substitutions, it is not clear whether there exist any substitutions satisfying this constraint, or whether such substitutions are unique. The following theorem states that in fact, given our assumptions, there is essentially a unique implementation.

THEOREM 1. *If $P$ is a joint atemporal knowledge-based protocol for environment $\mathcal{M}_e$, then there exists a substitution $\theta$ such that $skell(P)\theta$ is an implementation of $P$ in $\mathcal{M}_e$ with respect to* clk. *Moreover, for all substitutions $\theta, \theta'$ such that $skell(P)\theta$ and $skell(P)\theta'$ are implementations of $P$ in $\mathcal{M}_e$ with respect to* clk, *we have that $\mathcal{I}^{\text{clk}}(M(\mathcal{M}_e, P\theta)) \models X^t(\theta(v_\phi^t) \Leftrightarrow \theta'(v_\phi^t))$ for all $v_\phi^t \in skellVar(P)$.*

The result is similar to a result of [11]. Note that although $\theta(v_\phi^t)$ and $\theta'(v_\phi^t)$ may be distinct formulas, they are equivalent, in the context of any implementation, at the time of their relevance to the behaviour of the implementation. It follows that the systems $\mathcal{I}^{\text{clk}}(M(\mathcal{M}_e, P\theta))$ and $\mathcal{I}^{\text{clk}}(M(\mathcal{M}_e, P\theta'))$ are isomorphic with respect to the variables of $\mathcal{M}_e$ and $P$.

We now consider the synchronous perfect recall semantics. In this case, an agent's knowledge is semantically defined using not just the agent's current observation, but also using its past observations. Implementations of knowledge-based programs with respect to this semantics are therefore permitted to refer to these past observations. To enable this, we first introduce some new "history" variables to represent the past observations, and then state the perfect recall semantics as an application of the clock semantics.

Given a joint knowledge-based program $P$ of length $m$, let $P^h$ be the knowledge based program obtained after making the following modifications to $P$:

1. if $OVar_i$ is the set of observable variables for agent $i$, replace this by the set $OVar_i^h = OVar_i \cup \{v@k \mid v \in OVar_i, 0 \leq k < m\}$;

2. replace $LVar_i$ by $LVar_i \cup \{v@k \mid v \in OVar_i, 0 \leq k < m\}$;

3. replace each atomic statement $\ll a \mid \alpha \gg$ at time $k$ in $\text{Prog}_i$ by the statement $\ll a \mid \alpha, \beta \gg$, where $\beta$ is the collection of assignments $v@k := v$ for $v \in OVar_i$.[3]

Intuitively, each variable $v@k$ is a new local observable variable that records the value of the original observable variable $v$ of agent $i$ at time $k$. We now define an implementation of $P$ in $\mathcal{M}_e$ with respect to the synchronous perfect recall semantics to be an implementation of $P^h$ in $\mathcal{M}_e$ with respect to the clock semantics. By Theorem 1, such implementations are also guaranteed to exist and are behaviourally unique.

## 4. SYNTHESIS

The semantics for knowledge-based programs requires that the (semantically unique) implementing substitution $\theta$ for all knowledge conditions be given, and then verified for correctness. We now describe an incremental construction of this substitution that serves as the basis for our symbolic synthesis procedure. For the remainder of this section we fix an environment model $\mathcal{M}_e$ and a joint knowledge-based program $P$. Let $N$ be the maximal time of

---

[3]Our implementation optimizes this by sharing history of observed environment variables between agents.

(a)  **if** $K_i muddy_i \vee K_i \neg muddy_i \rightarrow \ll \text{SayYes} \mid \gg$
     $[] \neg(K_i muddy_i \vee K_i \neg muddy_i) \rightarrow \ll \text{SayNo} \mid \gg$ **fi**

(b)  **if** $v^0_{K_i muddy_i \vee K_i \neg muddy_i} \rightarrow \ll \text{SayYes} \mid \gg$
     $[] \; v^0_{\neg(K_i muddy_i \vee K_i \neg muddy_i)} \rightarrow \ll \text{SayNo} \mid \gg$ **fi**

**Figure 2: A knowledge-based program (a) and its skeleton (b)**

occurrence of any knowledge condition in P. Let $skell(\text{Prog}_i) = stat^i_1; \ldots stat^i_m$.

For the clock semantics, we work with epistemic Kripke structures $M(S) = (S, \{O_i\}, \pi)$, where $S$ is a set of assignments to $Var_e \cup \bigcup_{i \in Ags} LVar_i$, the observation functions are just restrictions to the observable variables, i.e., $O_i(s) = s \upharpoonright OVar_i$, and $\pi$ is just the trivial interpretation on $S$, i.e., $v \in \pi(s)$ iff $s(v) = 1$.

In particular, for $k = 0 \ldots N$ we define structures $M_k = M(S_k)$ by defining the sets $S_k$. At the same time, we define the substitution $\theta$. These definitions proceed inductively, as follows. First, we define $S_0$ to be the set of assignments $s$ such that $s \models Init_e \wedge \bigwedge_{i \in Ags} Init_i$. This determines $M_0$.

Assuming that $M_k$ has been constructed, we next define the implementation $\theta(v^k_\phi)$ of each knowledge condition $\phi$ in $\text{Prog}_i$ at time $k$. This implementation is required to be a boolean expression over the set $OVar_i$ of observable variables for agent $i$. Rather than give this formula explicitly, we characterize it by describing the assignments $o$ to these variables on which the formula is satisfied. For $v^k_\phi \in skellVar(\text{Prog}_i)$, we let $\theta(v^k_\phi)$ be any formula such that for all $o = O_i(s)$ with $s \in S_k$, we have $o \models \theta(v^k_\phi)$ iff $M_k, o \models_i \phi$. (This does not necessarily uniquely define $\theta(v^k_\phi)$ on all possible observations, but leaves some flexibility to optimize the size of the formula by choosing its value appropriately on the "don't-care" observations, applying ideas familiar from digital circuit design theory [21].)

Next, we define

$$S_{k+1} = \{t \mid \exists s \in S_k ((s, \tau \|_{i \in Ags} stat^i_k \theta) \longrightarrow (t, \tau \|_{i \in Ags} \epsilon))\} \;.$$

That is, we run the $k$-th step of the knowledge-based programs using the implementations of the knowledge conditions as just defined from $M_k$, using the operational semantics $\longrightarrow$ for standard programs. (Note that the substitution $\theta$ has not yet been completely defined, but it has already been sufficiently defined to provide a value for each $v^k_\phi$ in $stat^i_k$, so that $stat^i_k \theta$ is a standard program not containing any skeleton variables $v^j_\psi$.) This now gives the structure $M_{k+1} = M(S_{k+1})$.

The following result states that the substitution obtained by this process provides an implementation of P.

THEOREM 2. *Let $\theta$ be the substitution defined above. Then $P\theta$ implements P in $\mathcal{M}_e$ with respect to the view* clk.

The iteration using the epistemic structures $M_k$ in this construction is a generalization of an algorithm already in use in MCK for model checking standard programs with respect to specifications of the form $X^k \phi$, with $\phi$ an atemporal formula, interpreted with respect to the clock semantics. In the case of standard programs, the substitution $\theta$ is the empty substitution, and the construction simplifies to the existing algorithm in that case. The existing algorithm was already implemented symbolically using BDD's (see section 2.1) to represent the structures $M_k$, and the implementation is easily generalized to cover the extensions above. The main change is that it is now required at each stage to evaluate the applicable knowledge formulas $\phi$ in the structures $M_k$. This is done using an existing algorithm that computes a BDD representation of the set of states of $M_k$ satisfying $\phi$, given the BDD representation of $M_k$. The concrete condition $\theta(v^k_\phi)$ is then extracted as a boolean expression over observable variables that holds in $M_k$ at the same assignments to observable variables as the formula $\phi$. (Note that since $\phi$ is a boolean combination of observable variables and formulas $K_i \psi$, its satisfaction depends only on observable variables.)

Since the semantics of knowledge-based programs with respect to the synchronous perfect recall semantics has been introduced above by means of a reduction to the clock case, we note that we also obtain a procedure for synthesis of implementations with respect to the synchronous perfect recall semantics. The only change required is the introduction of history variables as described above.

## 5. EXAMPLES

In this section we discuss the performance of the symbolic synthesis approach on a number of simple examples, and compare it to an explicit state approach to a closely related problem.

The explicit state approach is essentially that implemented in DEMO [37], which is the only other model epistemic checker that presently has the expressive power to handle a problem close to the knowledge-based program synthesis problem that our system is able to handle. However, compared to our formulation, DEMO does not include knowledge-based programs as an explicit construct, it does not attempt to synthesize a concrete implementation of such a programs, and it can handle only situations where the atomic propositions do not change value over time.

DEMO deals with the evaluation of statements $(M, S) \models [U, T]\phi$, where $M$ is an epistemic model, $S$ is a set of states of that model, $U$ is an epistemic update and $T$ is a set of states of $U$. More precisely, $M = (W, \{\sim_i\}_{i \in Ags}, \pi)$ where $W$ is a set of worlds, each $\sim_i$ is an equivalence relation on $W$, and $\pi : W \rightarrow Prop$ is an interpretation of the atomic propositions. The update structure $U$ has the form $(E, \{\sim^E_i\}_{i \in Ags}, pre)$, where $E$ is a set of *events*, each $\sim^E_i$ is an equivalence relation on $E$ representing events that agent $i$ is not able to distinguish, and $pre$ maps $E$ to formulas (since it is all we will need, we assume here that these formulas are atemporal but possibly epistemic formulas in our language). Intuitively, $pre(e)$ is a pre-condition for the occurrence of event $e$. The *update of $M$ by $U$* is then defined to be the epistemic structure $M \circ U = (W', \{\sim'_i\}, \pi')$ where $W' = \{(w, e) \in W \times E \mid M, w \models pre(e)\}$, the relation $\sim'_i$ is defined by $(w_1, e_1) \sim'_i (w_2, e_2)$ if $w_1 \sim_i w_2$ and $e_1 \sim^E_i e_2$, and $\pi'(w, e) = \pi(w)$. The statement $(M, S) \models [U, T]\phi$, where $S \subseteq W$ and $T \subseteq E$, holds just when $M \circ U, (w, e) \models \phi$ for all $w \in S$ and $e \in T$ with $w \models pre(e)$.

In the special case where actions do not change the values of propositions (one example where this holds is the Muddy Children problem, discussed below) we can encode each stage of a knowledge-based program as an update. Suppose that each agent $i = 1 \ldots n$ has atomic statement

$$\textbf{if } g^i_1 \rightarrow a^i_1 \; [] \ldots [] \; g^i_{k_i} \rightarrow a^i_{k_i} \textbf{ fi}$$

where the $g^i_j$ are atemporal epistemic formulas. Then the parallel composition of these statements corresponds to a set $E = \Pi^n_{i=1}\{1 \ldots k_i\}$ with $pre(j_1, \ldots j_n) = \bigwedge_{i=1 \ldots n} g^i_{j_i}$. If the effect of the actions $a^i_j$ on observable variables (as described in the environment model) can be captured by indistinguishability relations on $E$, then we can encode the stage of the knowledge-based program as an update.

One difference is immediately apparent however: in DEMO, the epistemic model $M$, the update structure $U$, and the structure $M \circ U$ are all represented by an explicit enumeration of their states. Be-

```
muddy: Bool[Agent]
info: Bool[Agent]

init_cond = (Exists x:Agent() (muddy[x])) /\ Forall x:Agent() (info[x] == muddy[x])

agent Child0 "child" ( info[Child1], info[Child2], info[Child3] )
agent Child1 "child" ( info[Child2], info[Child3], info[Child0] )
agent Child2 "child" ( info[Child3], info[Child0], info[Child1] )
agent Child3 "child" ( info[Child0], info[Child1], info[Child2] )

transitions
begin
info[Child0] := Child0.SayYes; info[Child1] := Child1.SayYes;
info[Child2] := Child2.SayYes; info[Child3] := Child3.SayYes
end

protocol "child" ( info1: observable Bool, info2: observable Bool, info3: observable Bool )
begin
 if (Knows Self muddy[Self]) \/ (Knows Self neg muddy[Self]) -> << SayYes >>
 [] otherwise -> skip  fi;
 if (Knows Self muddy[Self]) \/ (Knows Self neg muddy[Self]) -> << SayYes >>
 [] otherwise -> skip  fi;
 if (Knows Self muddy[Self]) \/ (Knows Self neg muddy[Self]) -> << SayYes >>
 [] otherwise -> skip  fi;
 if (Knows Self muddy[Self]) \/ (Knows Self neg muddy[Self]) -> << SayYes >>
 [] otherwise -> skip  fi
end
```

**Figure 3: A knowledge-based program for muddy children (perfect recall version)**

cause of the cartesian products in the definition, the size of these state spaces potentially grows exponentially in the number of agents (and the number of updates applied to an initial structure), although DEMO applies a quotient under a maximal bisimulation that may reduce the size of these spaces in some cases. Our symbolic representation, on the other hand, has the potential to avoid this exponential blow-up. (The benefit is only potential because BDD representations, though they often prove to be small in practice, are also not guaranteed to be small in all cases.) It is therefore interesting to investigate whether this potential benefit is realized in interesting cases. We now explore this question for the well-known Muddy children problem.

### 5.1 Muddy children

The muddy children puzzle [10] can be stated as follows:

> A group of *n* children have been playing outside, and some have mud on their foreheads. Each child can see the forehead of the others but cannot see his or her own forehead. Father says to group, "At least one of you has mud on your forehead". He then repeated asks following question: "Do you know whether or not you have mud on your forehead?" The children give their answers ('Yes" or "No" ) simultaneously each time the question is asked, and each child observes the answers given by the other children.

Assuming that the children are perfect reasoners, have perfect recall and are honest, the expected behaviour is that if *k* out of *n* of the children are muddy, then all children will answer "No" until round *k*, in which all the muddy children answer "Yes" and the clean children answer "No". We note also that from round *k* + 1 all children answer "Yes".

The puzzle can be represented as a knowledge-based program. Figure 3 gives the representation of the environment and the children's protocol in the concrete syntax of our MCK implementation, for the perfect recall case with *n* = 4. Father's statement is captured by means of the statement init_cond, which defines the set of initial states. Since these are common knowledge, it is initially common knowledge that there is at least one muddy child. (The existential/universal quantifiers are restricted to finite types and are just a syntactic sugar for disjunction/conjunction.) An agent's observable variables $OVar_i$ are declared using the keyword observable. Each observable variable adds complexity to the BDD computation and, in the perfect recall semantics, is moreover replicated at each moment of time. To minimize these costs, we use a variable info[x] that represents the new information concerning agent *x* at each step. Initially this variable is used to represent whether agent *x* is muddy, and at later steps it represents whether agent *x* has just said "Yes". Each agent observes all the variables info[x] for the *other* agents (it can always deduce whether it would itself have said "Yes" at the previous step).

Symmetry of the children's behaviour is handled by giving a general description, the knowledge-based protocol "child". The knowledge-based program consists of a repeated sequence of if statements, in which the keyword otherwise represents the negation of the disjunction of all the preceding guards. The agent statements create fresh instances of the general protocol, in which the the parameters of the instance are aliased to the corresponding environment variables, and the keyword Self in the protocol is interpreted as the agent being defined. In particular, each such statement says that a child can observe (via the observable variables $see_i$) the new information about the other children. The environment's transitions clause simply stores the childrens' answers to the boolean variables info[x] for $x \in Ags$.

The representation we use for a clock semantics version is slightly different. Here, we cannot rely on agents to remember whether they initially observed other children to be muddy, or whether they said "Yes" in the previous step. We therefore make observable to all agents an array said, representing the previous statements of all the agents, and also include an observable variable for the muddiness of each other child. Interpreting this version with respect to the clock semantics yields exactly the same behaviour of the children in the implementation as the perfect recall version, viz., if there are *k* muddy children then these children first say "Yes" at stage *k*.

To represent the puzzle in DEMO, each step of the knowledge based program needs to be represented as an update structure $U = (E, \{\sim_i\}_{i \in Ags}, pre)$, in which the events $E$ correspond to possible observations that are made by the children after they reply to father's question. Thus, for $n$ agents, the set of events $E = \{0, 1\}^n$, with $e \sim_i e'$ iff $e = e'$, and for each tuple $e = (e_1, \ldots, e_n) \in E$ we have that $pre(e) = \bigwedge_{i=1}^n \phi_i$ where $\phi_i = (K_i \mathtt{muddy[i]}) \vee (K_i \neg \mathtt{muddy[i]})$ if $e_i = 1$ and $\phi_i = \neg((K_i \mathtt{muddy[i]}) \vee (K_i \neg \mathtt{muddy[i]}))$ if $e_i = 0$. Since DEMO runs in the Haskell interpreter, it is possible to represent $U$ succinctly as a Haskell program, but to perform the update calculation DEMO needs to construct the set $E$ explicitly, so necessarily performs an exponential amount of work.

The experimental results[4] comparing the performance of our symbolic approach to the DEMO modelling are shown in Table 1. In addition to synthesis, we check, for the MCK program, the formula $X^n \phi$ where $\phi = \bigwedge_{i \in Ags}(K_i \mathtt{muddy[i]} \vee K_i \neg \mathtt{muddy[i]})$ which expresses that after $n$ rounds, all the children know whether they are muddy. For the DEMO program, we check $\phi$ after updating $n$ times.

Note that for $n$ muddy children, we are dealing with an initial state space of $2^n - 1$ states and a deterministic solution protocol that runs for $n$ steps, giving $n \cdot (2^n - 1)$ points in the relevant part of the interpreted system. The results demonstrate that our symbolic approach does in practice scale significantly better when dealing with this exponentially growing problem, particularly in the case of the clock semantics version. (Recall that the synthesized behaviour is exactly the same as for the perfect recall interpretation.) DEMO's performance rapidly degrades as the number of agents increases, whereas our symbolic approach to the clock semantics is able to very efficiently handle problems of larger scale. On the other hand, DEMO is implicitly computing the perfect recall solution, so an arguably the fairer comparison is with the MCK perfect recall model. Here too our approach scales better, e.g., handling 10 agents in time comparable to DEMO's time for 7 agents. However, after initially lagging DEMO's explicit state approach, the total running time for the larger cases that can be handled becomes more than two orders of magnitude better.

## 5.2 Leader Election

The second example we consider concerns maintaining knowledge of the leader on a ring of agents. We suppose that there are $n$ agents numbered $i = 1 \ldots n$, with agent $i$ able to send messages to agent $(i \mod n) + 1$. Each agent has an observable input buffer that is able to store one message. The agent also observes its own agent number. An agent can crash at any time, and once crashed, remains crashed. The leader at time $t$ is defined as the highest numbered agent that has not crashed by time $t$ (and otherwise 0).

In each round, the environment first crashes a subset of the agents. All uncrashed agents may send a message. The network delivers any message that an agent is trying to send to the intended recipient, provided that the sender has not crashed. If agent $i$ has crashed, then the network detects this, and in place of the message that agent $i$ would have sent, the network delivers the message that was in agent $i$'s buffer to agent $(i \mod n) + 1$. (Intuitively, if the network cannot deliver a message to an agent it sends it to the next agent in the ring.) Each message is reliably marked with a "from" field, so that the recipient can determine the original sender of the message. (Note that this means also that when it receives a message that is marked as from an agent other than the ("lower" numbered) agent to its left, an agent can deduce that all agents between itself and the

---

[4] Our experiments were conducted on a Ubuntu Linux system (3.06GHz Intel Core i3 with 4G memory). Each process is allocated up to 500M memory.

original sender of the message have crashed.)

Note that the definition of the leader is a global property. Since it takes at least $n$ rounds of communication for any information about a node to reach all other nodes, if the leader crashes then another distant non-crashed agent cannot know about the crash for several steps: agents distant from the leader therefore cannot know whether the actual leader is still alive, so cannot know who is the leader. We therefore focus on a weaker property than knowledge of who is the leader. We say that agent $i$'s presumed leader is the largest agent number $m_i$ for which agent $i$ considers it possible that $m_i$ is the leader. To help acquire and spread this knowledge, the agents inform each other about their presumed leader: in each round, each (noncrashed) agent $i$ sends its neighbour a message "*from i: j*" stating that its presumed leader is $j$.

Figure 4 shows our MCK representation of the knowledge-based protocol. (For space reasons we omit the program for the environment.) To help identify crashes in the first step, we set the message in agent $i$'s buffer at time 0 to be *from i: 0*. The atomic statement $\ll$ Sendj | presumed := j $\gg$, performed by agent $i$, has the effect (encoded in the program for the environment) of causing the message "*from i: j*" to be delivered to the right neighbour of agent $i$ provided that the agent has not crashed. The assignment to local variable presumed stores the current presumed value.

It can be seen that the state space for this problem grows rapidly: since a run is determined by the time at which each agent crashes, if at all, for $n$ agents there are $(k + 1)^n$ runs of length $k$. Table 2 shows the performance of our symbolic synthesis procedure as we increase the number of steps of the protocol. (We do not give a comparison to DEMO. This problem is beyond the scope of DEMO, because it handles only static propositions, whereas in this problem the propositions change value over time.)

We have confirmed by model checking a manual solution for the 3 agent case that that under both the clock and perfect recall semantics, at each step, an agent knows that the leader is not $A3$ just when one of the following holds:

1. it knew this already in the previous step, because it already had presumed $< 3$,

2. it receives a message from another agent that must have passed through a chain of failures including A3, or

3. it receives a message "from $j$: $y$" with $y < 3$, which implies that agent $j$ knows that $A3$ is not the leader, or

4. it receives the message "from 3: 0", which implies that A3 failed in the first step.

Essentially the same predicate (with 2 in place of 3) captures the circumstances under which an agent knows that the leader is not 2, provided it also knows that the leader is not 3.

Manually verifying a nonterminating protocol that uses the above predicates at each step to determine the current presumed leader, we can verify that the following properties holds in the resulting protocol (for the case of 3 agents): at all times all noncrashed agents are greater than or equal to the actual leader, and if there are no more crashes after time $t$, then within 2 steps all non-crashed agent's presumed leaders are the same as the actual leader.

## 6. RELATED WORK

Our focus in this paper has been on the pragmatics of knowledge-based program syntax and on synthesis using a particular data structure for the symbolic representation of knowledge-based program implementations. A number of works have approached the problem

```
protocol "elect" (crashed : Bool,  my_num: observable LeaderNum,
                 from_field: observable LeaderNum, message: observable LeaderNum)

presumed: LeaderNum

init_cond = presumed == 3

begin
if  (neg crashed) /\ neg Knows  Self neg leader == 3 ->  <<Send3 | presumed := 3 >>
[] (neg crashed) /\ (Knows  Self neg leader == 3 )
                 /\ neg Knows  Self neg leader == 2 ->  <<Send2 | presumed := 2 >>
[] (neg crashed) /\ (Knows  Self neg leader == 3 )
                 /\ (Knows   Self neg leader == 2 )
                 /\  neg Knows  Self neg leader == 1 -> <<Send1 | presumed := 1 >>
[] otherwise -> skip fi;
(repeat if statement)
end
```

**Figure 4: A knowledge-based protocol for leader election**

| No. of Children | 4 | 5 | 6 | 7 | 8 | 9 | 10 |
|---|---|---|---|---|---|---|---|
| DEMO | 0.54 | 5.79 | 71.11 | 897.28 | 9,995.10 | > 36,000 | |
| MCK clk | 0.32 | 0.88 | 2.03 | 6.32 | 9.09 | 20.23 | 57.30 |
| MCK spr | 1.14 | 5.96 | 13.13 | 58.92 | 96.12 | 484.22 | 1,239.60 |

**Table 1: Running Times (seconds) of Muddy Children**

| No. of Agents, Semantics | Length of Run | | | | | | | | |
|---|---|---|---|---|---|---|---|---|---|
| | 2 | 3 | 4 | 5 | 6 | 7 | 8 | 9 | 10 |
| 3, MCK clk | 2.35 | 4.42 | 8.58 | 9.95 | 21.42 | 29.56 | 28.39 | 33.92 | 35.58 |
| 3, MCK spr | 11.26 | 10.43 | 63.01 | 170.16 | 1,607.82 | 5,971.44 | 26,624.59 | > 36,000 | |

**Table 2: Performance of synthesis on election protocol (seconds)**

of constructing implementations from a more theoretical perspective.

Besides identifying the synchronous atemporal case, that we have treated here, as one in which unique implementations exist, in [11] it is shown that deciding the existence of an implementation with respect to the observational view in a finite state environment is PSPACE complete, even when the knowledge conditions are expressed using linear time temporal logic operators. Since model checking LTL is also PSPACE complete but is still considered practical, this might suggest that this knowledge-based program implementation problem should also be tractable; unfortunately the algorithm in question requires guessing an implementation from an exponentially large set and then verifying it, so it is not clear that this is the case.

We have focussed on programs of bounded length. It is shown in [32] that the problem of determining whether an atemporal formula of the form $K_i\phi$ where $\phi$ is propositional, holds at a given view of length $n$ in the implementation of a knowledge-based program with respect to the synchronous perfect recall view can be as hard as PSPACE-complete. Besides indicating that we cannot expect to always obtain tractable implementations in the perfect recall case even for programs of bounded length, this result also has implications for nonterminating knowledge-based programs: it implies that implementations of such programs are not finite state encodable in general. However, this does not preclude the practicality of synthesis in particular cases.

For example, it is shown in [31] that finite state implementations of nonterminating knowledge-based programs are guaranteed to exist in the case of the clock view, as well as broadcast environments and environments with a single agent with synchronous perfect recall. A formal verification of these results is described in [12]. The implementation approach we have considered in the present paper can in principle be extended to construct such implementations, but we have not yet experimented with this.

A general scheme that constructs a finite state implementation with respect to the perfect recall semantics in the (undecidable) situation that one exists is described in [33]. The construction exploits a quotient by the maximal bisimulation on temporal slices, that is similar to the optimization used in the DEMO implementation. We refer to Section 5 for a comparison of our approach to the epistemic update logic problem considered in DEMO.

A number of papers have also applied model checking of knowledge properties to synthesize distributed control strategies [4, 14, 22]. However, the approach taken in these works is weaker than that in knowledge-based programs. Roughly, it corresponds to taking just one iteration of the fixpoint operator for a knowledge-based program, so that it is not guaranteed that the implementing condition is equivalent to the desired knowledge property in the protocol synthesized.

## 7. CONCLUSION

Our contribution in this paper has been to take the first step towards the goal of a practical tool, based on symbolic methods, for knowledge-based program implementation. We have demonstrated that the approach works on two modest scale examples. In future work, we plan to undertake further application case studies. We also intend to develop optimizations of our initial implementation: we believe that many avenues remain open for improvement of the performance of our system. We also plan to extend it in directions such as handling non-termination and probabilistic knowledge.